\documentstyle{aa}     % LaTeX A&A  Standard Fonts

\begin{document}

   \title{The peculiar B-type supergiant HD327083}\thanks{Based on observations 
made at Observatorio do Pico dos Dias, operated by MCT/Laboratorio Nacional de 
Astrofisica, Brazil}
 
   \author{M.A.D. Machado\inst{1}, F.X. de Ara\'ujo\inst{2}, S. 
Lorenz-Martins\inst{1} }

   \institute{Observatorio do Valongo/UFRJ,
              Ladeira Pedro Antonio 43,
              20080-090 - Rio de Janeiro - Brazil
   \and
              Observatorio Nacional/MCT, 
              Rua Gal. Jose Cristino 77,
              20921-400 - Rio de Janeiro - Brazil}
 
   \date{Received  / Accepted }

   \offprints{F.X.Araujo (araujo@on.br)}
   
 \input epsf

\abstract{ 
Coud\'e spectroscopic data of a poorly-studied peculiar
supergiant, HD327083, are presented. H$\alpha$ and H$\beta$ line 
profiles have been fitted employing a  non-LTE code adequate for 
spherically expanding atmospheres. Line fits lead to estimates
of physical parameters. These parameters suggest that HD327083 may be
close to the Luminous Blue Variable phase but it is also possible that it
could be a B[e] Supergiant.  
 }  

 \maketitle
 
\section{Introduction}

Luminous emission-line B-type stars often present massive winds. 
Many of them share some common properties: i) optical spectra dominated by
HI and FeII emission lines, usually displaying P-Cygni profiles;
ii) presence of forbidden  emission lines, 
specially of [FeII]; iii) strong red/infrared excess. Such properties are 
typical of the B[e] Supergiants (Lamers et al. 1998) but are also often seen
in the Luminous Blue Variables (LBVs). Their localization in the H-R diagram is
a real problem. The distances are very uncertain (particularly for 
galactic objects) and the absence of photospheric lines renders the 
assignment of a spectral type quite difficult.

The present paper concerns one of such interesting stars, namely HD327083. 
Almost no study exists in the literature devoted to it, but some works on 
emission-line objects have included it. H$\alpha$ emission was discovered
by Merrill \& Burwell (1949). Carlson \& Henize (1979) included HD327083 in 
their list of peculiar southern emission-line objects. They have proposed
a B8 spectral class. Some years ago Lopes et al. (1992) showed  
observations of some spectral regions. They revealed HI lines with P-Cygni
profiles as well as FeII $\lambda\lambda$ 4924, 5018, 5169 \AA \, (m42). On
the  other hand the FeII $\lambda$ transition at 9997 \AA\, seemed to be in
pure emission. The authors  have suggested a B6 type. More recently, Th\'e et
al. (1994) presented a new  catalogue of (confirmed and candidates) Herbig AeBe
stars. In this catalogue HD327083 is cited among the extreme emission-line
objects, which are not  identified to belong to a certain group. However, they
proposed the spectral type B1.5. The same assignment has been used by Valenti
et al. (2000) while Sheikina et al. (1999) have obtained log $T_{\rm eff}$
$\geq$ 4.2 and log ${L \over  L_\odot}$ $\sim$ 6.0. 

Here we show our recent spectroscopic data and estimates of some physical
parameters (as $\dot M$,  $L_{\star}$, $T_{\star}$ and $A_{\rm He}$, the 
numerical
abundance ratio $n$(He)/$n$(H)).
H$\alpha$ and H$\beta$ line profiles have been fitted with this aim.
These estimates allow us to locate  (within certain uncertainties) HD327083
in the H-R diagram. A discussion of its evolutionary status, based on the
derived  parameters, is also presented.  

\section{The data}

The spectra were obtained in August, 1999 using a Coud\'e spectrograph at 
the LNA 1.6m telescope in Pico dos Dias (Brazil). A grating of 600 l/mm
combined with a  1024$\times$1024 CCD have been used; the
reciprocal dispersion is about 0.4 \AA $\,$ pixel$^{-1}$ yielding a wavelength
range of $\sim$ 400 \AA  $\,$. The central wavelengths are: 
$\lambda_{\rm c}$ = 4750, 5000, 5200 and 6400 \AA. The data have been reduced 
using IRAF routines. We followed standard procedures:
bias subtraction, flat-fielding, linearization, wavelength calibration and 
correction for atmospheric extinction. 

\begin{figure}[t]
\epsfxsize=\hsize
\epsfbox{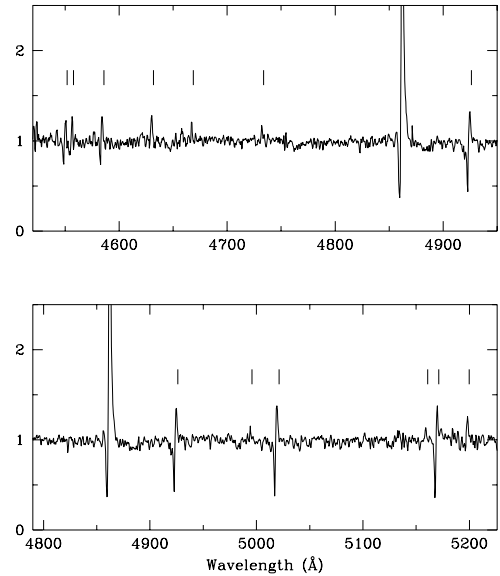}
\caption{Normalized spectra: $\lambda_{\rm c}$ = 4750, 5000 \AA\ (see text).}
\end{figure}

\begin{figure}[t]
\epsfxsize=\hsize
\epsfbox{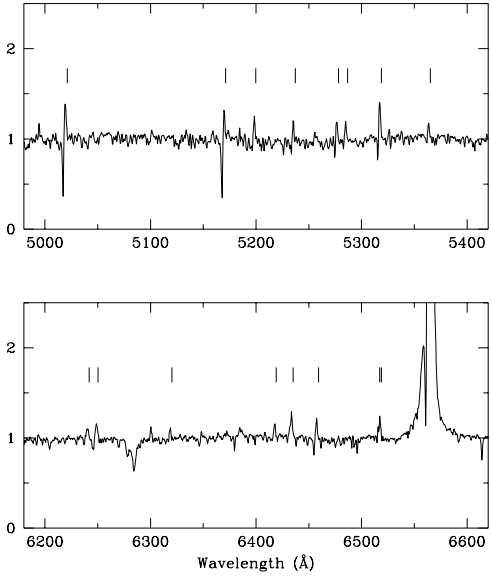}
\caption{Normalized spectra: $\lambda_{\rm c}$ = 5200, 6400 \AA\ (see text).}
\end{figure}

In Figs. 1-2 we can see our normalized spectra. In fact the spectral regions
covered by our observations are dominated - besides H$\alpha$ and H$\beta$
-, by FeII emission lines. Some of these lines show a P-Cygni profile and
others are in pure emission. Table 1 lists the FeII transitions identified,
which are marked in Figs. 1-2. Column 1 gives the measured wavelength (\AA) \,
of emission peaks, followed by our identification (Column 2).
We note that only one feature, at $\lambda$ 5159.0 \AA, has been recognized 
as a forbidden transition (see Fig. 1b). On the other hand [OI] $\lambda$  
6300.2 \AA \, is present (Fig. 2b). 

\begin{table}
\caption{FeII lines}
\begin{tabular}{|l|c|}\hline
Wavelength & Identification \\\hline
4550.0 & 4549.5 (m38) \\
4556.0 & 4555.8 (m37) \\
4584.0 & 4582.8 (m37) \\
       & 4583.8 (m38) \\
       & 4584.0 (m26) \\
4630.0 & 4629.4 (m37) \\
4667.0 & 4666.8 (m37) \\
4732.0 & 4731.4 (m43) \\
4924.5 & 4923.9 (m42) \\
4994.0 & 4993.4 (m36) \\
5019.5 & 5018.4 (m42) \\
5159.0 & 5158.0 [18F] \\
       & 5158.8 [19F] \\
5169.5 & 5169.0 (m42) \\
5198.0 & 5197.6 (m49) \\
5235.5 & 5234.6 (m49) \\
5276.5 & 5276.0 (m49) \\
5285.0 & 5284.1 (m41) \\
5317.0 & 5316.6 (m49) \\
       & 5318.8 (m48) \\
5363.5 & 5362.9 (m48) \\
6240.0 & 6238.2 (m74) \\
       & 6239.4 (m34) \\
6248.5 & 6247.6 (m74) \\
6318.5 & 6318.0($z^4D^o-c^4D$) \\
6417.5 & 6417.9 (m74) \\
6433.5 & 6432.7 (m40) \\
6457.5 & 6456.4 (m74) \\
6517.0 & 6518.2 (m72) \\\hline
\end{tabular}
\end{table}

\section{Summary of the model}

Let us briefly describe now the model employed.
Recently Machado (1998, 1999) developed a  non-LTE numerical code for
radiation transfer in expanding stellar atmospheres of massive stars.
This code is based on basic assumptions: spherical symmetry,
stationarity and homogeneity. The density structure is related to the
mass loss rate and the velocity field $V(r)$ via the equation of
continuity

\begin{equation}
\rho(r)={\dot M \over{4 \pi r^2 V(r)}}
\end{equation}

\noindent
with $\dot M$ as an input parameter and the velocity field is pre-specified
in  an ad-hoc way as a $\beta$-type law

\begin{equation}
V(r)=V_\infty(1-R_\star/r)^\beta
\end{equation}

\noindent
where $V_\infty$ is the terminal velocity of the wind and $\beta$ is
a measure of the rate of acceleration (de Koter et al. 1993).

The code is based on the following scheme. 
The statistical equilibrium equations are solved using the escape probability 
method for calculating the source function while the transfer equation is
solved using the ``SEI" - Sobolev Exact  Integration - method (Lamers et
al.1987). This is justified by the flow velocities observed in the wind of
HD327083.  

In the present version it is considered that the wind begins in the sonic
point localized at radius $R_\star$. The effective temperature $T_\star$ is 
defined by the luminosity $L_\star$ and the  radius $R_\star$ via the 
Stefan-Boltzmann law.

Regarding the temperature structure we have used the law suggested by Drew 
(1985):

\begin{equation}
T(r)= T_{\star} \left [ 0.78 - \left ( 0.51 {V(r) \over V_{\infty}} \right ) 
\right ]
\end{equation}

\noindent
We are aware that this temperature profile is more appropriate to main sequence 
O stars. Thus we have checked the influence of the temperature law. 
Some simple tests with isothermal and radially decreasing ad-hoc profiles  
have been performed. It was found that the final result is not sensitive
to similar temperature laws.  

The model atmosphere in the present work consists of hydrogen and helium.
For H and HeII we consider the lowest ten bound levels. For HeI we adopt the 
atomic model suggested by Almog \& Netzer (1989) that includes the individual 
levels until $n\leq$4 and combined levels with 5 $\leq$ $n$ $\leq$ 10. 

The determination of the population numbers of the energy levels is made  
in a non-LTE way solving the equations of statistical equilibrium for each 
level considered in the atomic model.The radiative transition probabilities  
and the photoionization cross section for all levels considered are quoted 
from the Opacity Project.

The ionization collision cross sections have been obtained from the approximate 
formulae given by Jefferies (1968). Excitation collisional rates have been  
calculated from the collisional strength values. For H and HeII these values 
are computed using polynomials given by Giovanardi et al. (1987). For HeI the 
collision strengths are taken from Berrington \& Kingston (1987) for
levels with $n\leq$4 and from Auer \& Mihalas (1968) and Mihalas \& Stone (1968)
to the other ones.

The radiation and the statistical equation systems are solved simultaneously 
through the wind that is divided in shells. A convergence
criterium is applied in order to move from one shell to the next.

\section{H$\alpha$ and H$\beta$ line fits}

\subsection{Methodology}
 
The  input parameters are the   temperature $T_\star$, the stellar radius 
$R_\star$ (and luminosity), the mass loss rate $\dot M$, the terminal velocity 
$V_\infty$, the helium abundance $A_{He}$ and the free parameter of the 
velocity 
profile $\beta$. In the present work we have fixed the value of the terminal 
velocity, $V_\infty$ = 400 kms$^{-1}$. This value is an average of 
the $V_{edges}$ of H$\alpha$ and H$\beta$ profiles. 

Due to the great number of ``free" parameters we have adopted the following 
strategy in order to obtain the line fits. First we defined temperature 
limits. Previous determinations of temperature for HD327083 are very uncertain.
As discussed in the introduction first estimates have pointed to values around
$T_{\rm eff} \, \sim$ 11000K - 13000K. More recent ones have lead to larger
values: $T_{\rm eff} \, \sim$ 16000K - 19000K. So we have decided to explore 
a broad range, that roughly corresponds to the LBVs and Yellow Supergiants 
domain: 8000K - 31000K. This $T_{\rm eff}$ range was fixed based on the
works by Gummersbach et al. (1995), Szeifert et al. (1996) and de Jager \&
Nieuwenhuijzen (1997). 

Secondly we have started computing models  according to the points of
the evolutionary tracks by Schaller et al. (1992) from  $M_{\rm ZAMS}$ = 
20 M$_{\odot}$ to 120 M$_{\odot}$. Each track consists of a number of points
which provide stellar temperature, luminosity, mass loss rate and surface 
abundance ratio ($A_{\rm He}$). These are the entry parameters 
for the code. The evolutionary points which give reasonably good fits 
(according to a $\chi^2$ criterium) are picked as starting points. From 
them we subsequently varied the input parameters, specially mass loss, 
in order to obtain better adjustments.

\subsection{Results} 

The mass loss rate is indeed small for the $M_{\rm ZAMS}$ = 20 M$_\odot$ and 
$M_{\rm ZAMS}$ = 25 M$_\odot$ paths. It is not enough to produce H$\alpha$
and H$\beta$ profiles with the intensities observed in HD327083. 
Concerning the $M_{\rm ZAMS}$ = 85 M$_\odot$ and the M$_{\rm ZAMS}$ = 120 
M$_\odot$ tracks, in such cases the He abundance is already high for stars 
just evolved off the Main Sequence. In these regions the temperatures
are high ($T_\star \geq$ 40000K) and appropriate fits of H$\alpha$ and
H$\beta$ cannot be achieved with reasonable values for the others  parameters. 

Our best fits have been obtained with the following set of parameters.
Model 1: $\dot M$ = 4.9 $\times$ 10$^{\rm -5}$ M$_{\odot}$ ${\rm yr^{-1}}$, 
$T_\star$ = 
19000K, $L_\star$ = 5$\times$ 10$^{\rm 5}$ L$_{\odot}$, $A_{\rm He}$ = 0.40.  
Model 2: $\dot M$ = 1.0 $\times$ 10$^{\rm -4}$ M$_{\odot}$ 
${\rm yr^{-1}}$, $T_\star$ = 9000K, $L_\star$ = 9.5$\times$ 10$^{\rm 5}$ 
$L_{\odot}$, $A_{\rm He}$ = 0.11. These fits are shown in Figs. 3a and 3b 
respectively and were both obtained with $\beta=3$, which 
indicates a slow expansion. Model 1 is based on the series of H$\alpha$ and
H$\beta$ profiles along the $M_{\rm ZAMS}$ = 40 M$_{\odot}$ path while 
model 2 is based on the $M_{\rm ZAMS}$ = 60 M$_{\odot}$ path. They have been
selected following the methodology described earlier. We note that the ``best
model" values of $\dot M$ are slightly higher (about 50\%) than those given 
directly
by the evolutionary paths.

\begin{figure}[t]
\epsfxsize=\hsize
\epsfbox{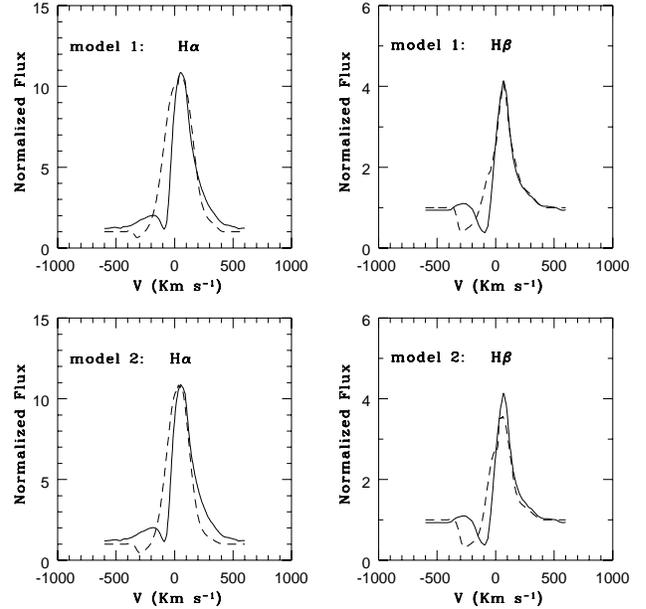}
\caption{Observational data (dashed lines) and best theoretical fits
(full lines) for H$\alpha$ and H$\beta$ profiles.}
\end{figure}

It can be seen that the agreement between observational and theoretical 
profiles is rather poor. The models reproduce well the emission strengths
but the absorption features are quite flat and narrow. This could suggest that
the emission originates in a high density disk. Unfortunately we cannot 
check this alternative within the scope of our spherical model. 

\section{Discussion}

Each set of parameters leads to a location in the H-R diagram. Fig. 4 
shows these positions and some evolutionary paths. Two alternatives 
- redward or blueward - are possible if only $T_\star$ and $L_\star$ 
are considered. He abundance may be invoked to choose one possibility
or another.

\begin{figure}
\epsfxsize=\hsize
\epsfbox{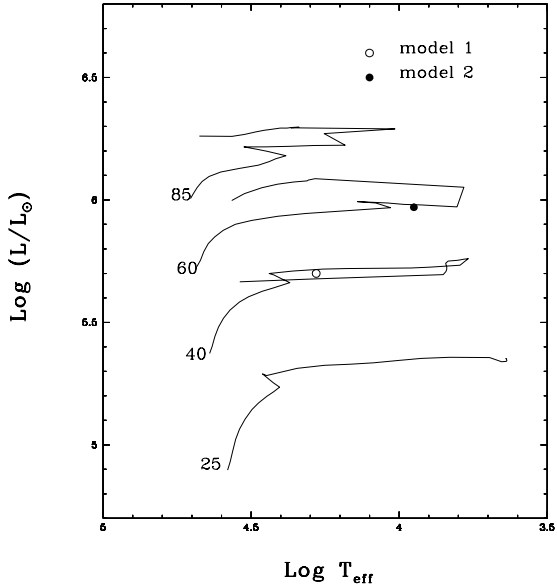}
\caption{Theoretical evolutionary tracks of 25 M$_{\odot}$; 40 
M$_{\odot}$; 60 M$_{\odot}$  and 85 M$_{\odot}$   
initial main-sequence mass.  The circles locate the best model parameters.}
\end{figure}

The evolutionary model 1 has $A_{\rm He}$ = 0.40.
This value is in the blueward
track: the star would be coming back to the hotter region of the diagram. 
The parameters in this scenario point to a LBV quiescent (or post-LBV) 
phase. On the other hand the evolutionary model 2 has $A_{\rm He}$ = 0.11.
This low value
indicates a point in the redward track. Here the 
star is in a less evolved stage, coming to the LBV eruptive phase.
The intensity of the He lines should, in principle, reflect the He
abundance. He lines are not seen in our data. We note however that our
spectra do not cover the wavelength regions with strong He lines,
such as $\lambda$ = 5876, 6678 \AA. So, we
prefer not to discard model 1 at present time. 

Both alternatives allow us to propose that HD327083 is near the LBV phase. 
However, this suggestion must be taken with caution. First because we 
have fitted only H$\alpha$ and H$\beta$ line profiles. So, the 
uncertainty in the derived parameters is large. Second because the high
value of $\beta$ needed to obtain ``good" fits reinforces the scenario 
of emission coming mostly from a flattened region, as the equatorial 
disk believed to be present in B[e] Supergiants (Zickgraf et al. 1986). 

In general the B[e] stars have stronger [FeII] lines than the LBVs. In our
spectra only the feature at $\lambda$ 5159.0 \AA\, (which we attribute to 
transitions in multiplets [18F] and [19F]) seems to be present in Fig. 1b, 
although it is uncertain in Fig. 2a. It is true that our data only covers 
the regions $\lambda \sim$ 4550 - 5400 \AA \, and  $\lambda \sim$ 6200 
- 6000 \AA. However, several forbidden transitions (from multiplets
[3F], [4F], [18F], [19F], [35F] etc) are in these regions. On the other hand
[OI] $\lambda$ 6300.2 \AA\, is clearly seen (Fig. 2b). This feature is intense
in the B[e] Supergiants but absent in the LBVs. 

It is likely that HD327083 is in a short-lived evolved stage of massive stars.
However it is still unclear if it is 
a LBV candidate or if it is a B[e] Supergiant. It is sure that it deserves to
be further investigated. Such analysis should be based on higher-quality
spectroscopic data, covering a broad spectral domain. Moreover it should 
include other information about the star such as the energy distribution, 
Balmer jump, and colour. From the theoretical point of view it is needed 
to employ a non-spherical code(e.g. Stee et al. 1995) to obtain adequate  
fittings not only for HI lines but also for HeI and N lines. This detailed
investigation will clarify the nature of HD327083 and it may contribute to a
better  understanding of evolved massive stars.

\section*{Acknowledgments}
%-----------------------------------------------------------------------------
The authors thank the questions and  comments of the referee, H. Lamers,
which greatly improved this paper.
M.Machado and S.Lorenz Martins acknowledge CNPq (300320/99-0) and FUJB (FUJB 
8635-5), respectively, for financial support.

\end{document}